# An integrated approach to simulating the vulnerable atherosclerotic plaque

Navid Mohammad Mirzaei,
Department of Mathematical Sciences,
University of Delaware,
DE 19716

Pak-Wing Fok,
Department of Mathematical Sciences,
University of Delaware,
DE 19716

and

William S. Weintraub,
Medstar Heart and Vascular Institute,
Washington DC 20010

**Abstract**

Analyses of individual atherosclerotic plaques are mostly descriptive, relying - for example – on histological classification by spectral analysis of ultrasound waves or staining and observing particular cellular components. Such passive methods have proved useful for characterizing the structure and vulnerability of plaques but have little quantitative predictive power.

In this viewpoint article, we propose an integrated quantitative framework to understand the evolution of plaque. The main approach is to use Partial Differential Equations (PDEs) with macrophages, necrotic cells, oxidized lipids, oxygen concentration and PDGF as primary variables coupled to a biomechanical model to describe vessel growth. The model is deterministic, providing mechanical, morphological, and histological characteristics of an atherosclerotic vessel at any desired future time point. We discuss the pros and cons of such a model, how such a model can provide insight to the clinician, and potentially guide therapy. Finally, we use our model to create computer-generated animations of a plaque evolution that are in qualitative agreement with serial ultrasound images published by Kubo et al. (2010) and hypothesize possible atherogenic mechanisms.

## Introduction

Cardiovascular disease affected more than 121 million people in 2016 in the United States and about 48% of adults (Benjamin, et al., 2019). It is often due to atherosclerosis, which manifests itself as a buildup of fatty deposits mainly in the intima of medium-sized and large arteries. Plaques exhibit considerable variability in their internal structure and histology. When they develop a thin cap and a large necrotic core, they are prone to mechanical rupture (Virmani, et al., 2003), which often results in thrombosis and acute events such as myocardial infarction (MI). The prediction of plaque progression and rupture remains one of the most important open problems in cardiovascular disease today.

Atherosclerosis has been studied from cellular (Bendeck, et al., 1994), genetic (Cyrus, et al., 1999) and clinical (Bourantas, et al., 2013) perspectives. Murine models have also proved valuable in elucidating the main aspects of early-stage atherosclerosis (Stadius, et al., 1992). However, one set of tools that remains underdeveloped is the application of deterministic mathematical models to aid in basic understanding of plaque characteristics. Historically, the analysis of plaque has been descriptive or statistical. Scientists may observe an individual plaque at a single time point or gather statistics from large cross-sectional studies. In either case, it is difficult to make predictions from observations at a single time point. In cross-sectional studies, individual risk-factors are assessed using statistical methods, but this approach may not be able to establish causal relationships. In this article, we advocate a paradigm shift to study the natural history of individual plaques using mathematical models. Combined with advances in imaging technology, we believe that such quantitative approaches will be pivotal in enhancing our understanding of the mechanisms of atherogenesis and plaque rupture. We believe that the best way to characterize the data being generated by imaging modalities such as ultrasound, optical coherence tomography (OCT) and palpography is by connecting them to and concurrently developing mathematical models.

In this article we will first review some paradigmatic models of atherosclerotic plaque that are popular in the engineering and mathematical communities. We then show how elements of these models can be combined into a biology-based computer-generated animation that illustrates plaque progression in terms of thickening of arterial layers, deformation of the vessel wall and changes in plaque histology.

**Plaque histology**

Plaque internal structure has been gradually elucidated since around the 1960s (Geer, et al., 1961). Researchers now believe that the internal structure of a plaque primarily determines its stability. Early plaques start life as "intimal cushions," "intimal thickenings" or "intimal xanthomas" (R, et al., 2002). These are relatively innocuous lesions. Over time, however, they can progress into "fatty streaks" which have a higher lipid content. Atheromas have regions of interior necrosis and are considered more dangerous since they are associated with cardiovascular events, e.g. stroke or myocardial infarction.

Enhanced ultrasound (virtual histology) protocols use machine learning to analyze the frequency content in ultrasound waves and classify atherosclerotic tissue into four different types: fibrous, fatty, necrotic and calcific (Nair, et al., 2001). The resulting patterns are fascinating and thought-provoking. For example, Kubo et al (2010) tried to gain insight into the dynamic evolution of plaque histology and morphology. Figure 1 shows serial IVUS images of plaque at baseline and after a 12 month follow-up. Data specific to individual plaques at these two time points were also collected. For example, morphological characteristics such as the area occupied by the lumen and fraction of the plaque occupied by the necrotic core were observed. This data set raises several important quantitative questions. How long does it take for a plaque to become vulnerable to rupture? How quickly do plaques grow in size? How quickly do regions of necrosis grow? How does necrosis affect the likelihood of rupture? All these questions can, in principle, be answered by a mathematical model of plaque development, properly calibrated against suitable data sets.

A popular modeling method in the applied mathematics community is to use Partial Differential Equations (PDEs). This approach describes how concentrations of certain cell types and metabolites change in time. For example, a simple system of PDEs that could describe the formation of hypoxia-induced plaque cores is:

$$\underbrace{\frac{\partial N}{\partial t}}_{\text{Rate of increase of NCs}} = \underbrace{D_1 \Delta N}_{\text{Diffusion of NCs}} + \underbrace{\gamma(C) M}_{\substack{\text{Generation of NCs} \\ \text{from dead} \\ \text{macrophages}}} - \underbrace{\beta_1 N}_{\substack{\text{Clearance of NCs by} \\ \text{other leukocytes}}} \quad (1)$$

$$\underbrace{\frac{\partial M}{\partial t}}_{\text{Rate of increase of MCs}} = \underbrace{D_2 \Delta M}_{\text{Diffusion of MCs}} - \underbrace{\nabla \cdot \left( \mu M \nabla (L + Q) \right)}_{\substack{\text{Chemotaxis to oxLDL} \\ \text{and CKs}}} - \underbrace{\gamma(C) M}_{\text{Hypoxic Death}} \quad (2)$$

$$\underbrace{\frac{\partial Q}{\partial t}}_{\substack{\text{Rate of increase}\\\text{of CKs}}} = \underbrace{D_3 \Delta Q}_{\substack{\text{Diffusion of}\\\text{CKs}}} - \underbrace{\beta_2 Q}_{\substack{\text{Natural decay}\\\text{of CKs}}} + \underbrace{\lambda_1 LM}_{\substack{\text{Production}\\\text{of CKs}}} \quad (3)$$

$$\underbrace{\frac{\partial C}{\partial t}}_{\substack{\text{Rate of increase}\\\text{of oxygen}}} = \underbrace{D_4 \Delta C}_{\substack{\text{Diffusion of}\\\text{oxygen}}} - \underbrace{\beta_3 C}_{\substack{\text{Background}\\\text{consumption}\\\text{of oxygen}}} - \underbrace{\lambda_2 CM}_{\substack{\text{Consumption of oxygen}\\\text{by macrophages}}} + \underbrace{f(x,y,t)}_{\substack{\text{Oxygen}\\\text{sources}}} \quad (4)$$

$$\underbrace{\frac{\partial P}{\partial t}}_{\substack{\text{Rate of increase}\\\text{of PDGF}}} = \underbrace{D_5 \Delta P}_{\substack{\text{Diffusion of}\\\text{PDGF}}} - \underbrace{\beta_4 P}_{\substack{\text{Natural decay}\\\text{of PDGF}}} \quad (5)$$

where $N$, $M$, $Q$, $C$, $L$ and $P$ represent concentrations of necrotic cells (NCs), macrophage cells (MCs), a chemokine (CK) such as Monocyte Chemoattracting Protein 1 (MCP1), molecular oxygen, oxidized Low Density Lipoproteins (oxLDLs) and Platelet Derived Growth Factor (PDGF); $D_1 - D_5$ are their diffusivities; $\mu$ is a chemotactic coefficient for macrophages; $\beta_1 - \beta_4$ are decay coefficients; $\lambda_1$ is a production rate of MCP1 and $\lambda_2$ is the consumption rate of oxygen by macrophages (see Table 1 for a summary). In our version of the model, we take the concentration of oxLDL ($L$) to be a known function in space and time.

In eq. (1), the term on the left hand side represents the time rate of change of necrotic cells at a given point **x** = *(x,y,z)* in the plaque. This equation specifies that the rate of change of NCs at **x** has three contributions. First, NCs from nearby points **x**+δ**x** can be randomly moved to **x**. The diffusion could arise from immune cells in the plaque exerting random forces on dead cells through their own random motion. Second, macrophage cells at **x** can die and become necrotic, essentially converting from vital cells to necrotic ones. The death rate γ depends on the oxygen concentration at that point, *C(x,y,z)* via the equation (Fok, 2012)

$$\gamma(C) = \gamma_{\min} + (\gamma_{\max} - \gamma_{\min})\left(\frac{C_{\text{crit}}^m}{C_{\text{crit}}^m + C^m}\right). \quad (6)$$

The normoxic death rate $\gamma_{\min}$ is the death rate for macrophages in an oxygen-sufficient environment and $\gamma_{\max}$ is the hypoxic death rate in an oxygen-limited environment. The last term in parentheses is called a Hill function and the number *m* is called the Hill coefficient (we take *m*=4). Eq. (6) states that macrophages die quickly when oxygen levels are low (C<$C_{\text{crit}}$), but slowly when levels are high (C>$C_{\text{crit}}$). We see that $C_{\text{crit}}$ acts as a hypoxic threshold for macrophage cells and the larger the value of *m*, the more abrupt the switch. Finally, NCs at that **x** can be cleared by leukocytes and the clearance rate is proportional to the number of NCs. The minus sign signifies that the clearance term *reduces* the rate of generation of NCs.

In eq. (2), the term on the left hand side represents the time rate of change of MCs. Again there are three contributions on the right hand side. Similar to eq. (1), the first term represents the random motion of the cells. The next term represents chemotaxis, the directed movement of cells towards chemoattractants such as oxLDL ($L$) and MCP1 ($Q$). Macrophages chemotax along the vector $-\nabla(L+Q)$ with associated "flux" $-\mu M \nabla(L+Q)$ and the coefficient µ capturing the speed of taxis. Note that $-\nabla(L+Q)$ is a vector that points from small values of *L+Q* to large values so MCs are modeled to move from low to high concentrations of total chemoattractant. The final term represents the death of macrophages with a death rate dependent on the local oxygen concentration $C$ as given by eq. (6). The $+\gamma(C)M$ in eq. (1) and $-\gamma(C)M$ in eq. (2) couple the equations together: when a macrophage cell dies, it converts to a necrotic cell.

In eq. (3), the term on the left hand side represents the time rate of change of CKs. Similarly, we have three terms on the right hand side. The first term corresponds to the random motion of the CKs. The next term represents the natural decay of CKs at a rate $\beta_2$. The last term represents the production of CKs by MCs after consumption of oxLDL at a rate $\lambda_1$. Motivated by the principle of mass action, the production rate is proportional to the product of macrophage and oxLDL concentrations, *LM*. The net production is greater if there are more MCs or oxLDLs particles.

In eq. (4), the term on the left represents the time rate of change of the oxygen concentration. The first term on the right hand side corresponds to the random motion of oxygen in tissue. The second term represents the background consumption of oxygen by all cells (excluding MCs) at position **x** with rate $\beta_3$. The third term represents the consumption of the oxygen concentration by macrophages at a rate $\lambda_2$. Again, motivated by mass action principles, the net consumption rate is proportional to the product of oxygen concentration and macrophage density, *CM*. The final term represents contributions from oxygen sources such as microvessels that could be present in advanced plaques. In eq. (5), the term on the left represents the time rate of change of the PDGF concentration. The first term on the right represents the random spread of PDGF. The second term corresponds to the natural (thermal) decay of PDGF at a rate $\beta_4$.

In principle, the solution of these equations produces *N*, *M*, *Q*, *C* and *P* as smooth functions of space and time, representing densities of NCs, MCs, CKs, O₂ and PDGF. When the governing equations are solved in practice, these quantities all find their equilibrium levels very quickly compared to the observed rate of plaque progression so the steady-state, time *independent* versions of (1)-(5) are actually computed by setting all time derivatives to zero. An example of the function *N* is shown in Figure 2, along with a thresholding method to formally distinguish between necrotic and fibrous tissue. To solve the equations, one needs the values of all the constants $D_1 - D_5$; $\beta_1 - \beta_4$; $\mu$, $\lambda_1$, $\lambda_2$, $\gamma_{\min}$, $\gamma_{\max}$, $C_{\text{crit}}$ and the Hill coefficient *m.* These constants can be found from experiments or estimated independently. The validity of these equations relies on the continuum assumption: rather than describing the behavior of individual cells, we are calculating their aggregate or average behavior using smooth functions. While this can be a limitation, there are more advanced methods that can be used to capture individual cell behavior (DiMilla, et al., 1991).

**Plaque Morphology and Mechanics**

In addition to describing the plaque's histology through PDEs, it is also important to describe the stresses within a plaque. (Virmani, et al., 2003) have shown that plaques are more likely to rupture if their caps are thin (<65 um). This observation becomes intuitive after we understand that caps in fibroatheromas rupture when they are stressed beyond a threshold yield stress (Loree, et al., 1992). As cap morphology evolves over time, so do the associated stress fields and the propensity for rupture. Therefore an integrated model of vulnerable plaque should also explain how stress and strain fields evolve and couple these fields to tissue growth and atrophy.

*Mechanics and Deformation.* Leonhard Euler (1707-1783) was possibly the first person to use mathematics to formulate models for arterial mechanics. His calculations described how the pulsatility of blood flow affected the expansion of arterial cross sections. By the 1970s arterial hemodynamics was a mature field and in the 1990s the mechanical properties of tissues became a core focus. Aided by developments in numerical methods for fluid mechanics and solid-fluid interactions, simulation of the cardiovascular system became a highly evolved enterprise: the ubiquity and power of computers allowed researchers to predict the deformations of the arterial wall and the hemodynamics contained within to an unprecedented level of detail. For example, (Simon, et al., 1993) developed a computational method based on the poroelasticity of arterial tissues. Their model coupled the wall deformation, fluid mechanics and associated transport phenomena in the arterial wall. (Auricchio, et al., 2001) investigated the biomechanical reaction of a stenotic artery wall to an expandable stent. Their aim was to understand the mechanisms underlying stent-related restenosis. Finally, (Akyildiz, et al., 2011) studied the effect of intima stiffness and plaque morphology on maximum plaque stress. They found that reducing cap thickness and increasing the size of the necrotic core increased the peak stress.

Many of the approaches outlined above utilize the *Finite Element Method* for computing the equilibrium configuration of an elastic artery under a given load. This is an elastostatics problem with no notion of time dependence. However, the framework can be extended to involve time using a model of tissue growth called morphoelasticity. The underlying assumption is that growth occurs so slowly that mechanical equilibrium is maintained at all times. Below, we set up the elastostatics problem, and then explain how morphoelasticity can be used to evolve the plaque dynamically.

*Elastostatics.* The deformation vector tracks the position of every material point in the artery under a given load and maps a *reference configuration* to a *deformed*

*configuration*. The starting point for our method is an energy integral that accounts for all the ways that deformations affect the total energy. This usually consists of three parts. First, there is the stored potential energy associated with deformations from an unstressed, reference state. Second, there is the energy associated with the deformations doing work against body forces (such as gravity). Finally the deformations also do work against surface forces (such as a lumen pressure). While deformations of the elastic body increase the energy, any work done against body and surface forces require energy and is subtracted from the total budget. In the absence of growth we have:

$$\Pi[\Phi] = \int_\Omega W(\nabla\Phi)\,dV - \int_\Omega f(\Phi)\,dV - \int_{\partial\Omega} g(\Phi,\nabla\Phi)\,dS, \quad (7)$$

where $\Pi$ is the total energy, $\Phi$ is the deformation vector, $f$ and $g$ are body and surface force densities, $\Omega$ is the "reference domain" (points in space occupied by the unloaded artery), and $\partial\Omega$ is the mathematical boundary between the lumen and the undeformed vessel wall. The strain energy density of the elastic artery is $W$ and depends on the deformation *gradient* $\nabla\Phi$. A quick way to understand this is that the energy should depend on the change in the dimensions of an infinitesimal cuboid relative to its dimensions in the reference configuration. In other words, the energy does not depend on the deformation vector $\Phi$ but rather on how $\Phi$ changes when applied to different points in the reference configuration. To find the arterial deformation, one first defines the mechanical properties of the vessel wall by specifying $W$. For example, it may be mechanically anisotropic (collagen fibers in the vessel make it harder to stretch radially, than axially); or it may be layer-dependent (mechanical properties of the intima are different than the media or adventitia). We employ a layer-specific strain energy following (Holzapfel, et al., 2005) which comes from *ex-vivo* stress tests of arterial tissue.

For most arterial problems, body forces such as gravity are negligible so $f = 0$. For a lumen pressure $p$, one can show that the surface force density is in fact

$$g(\Phi,\nabla\Phi) = -\frac{p}{3}J(\nabla\Phi)^{-T}N \cdot \Phi \quad (8)$$

where $J = \det \nabla\Phi$ and $N$ is the unit outward normal vector to $\partial\Omega$. The ultimate objective is to find the deformation $\Phi$ that minimizes $\Pi$ in (7). This high-dimensional optimization problem is solved using a computer. Once $\Phi$ is known, the displacement of every point in the reference configuration determines the deformed configuration: see Fig. 3. More details can be found in textbooks such as (Holzapfel, 2000).

*Tissue Growth and Morphoelasticity.* The way we account for tissue growth in our model is by employing a theory called morphoelasticity (Goriely, 2017). Eq. (7) above depends on the deformation gradient $\mathbf{F} = \nabla\Phi$ which captures how an infinitesimal cuboid of tissue changes dimensions under a deformation $\Phi$.

However, since biological tissue mostly consists of water, the volume of the cuboid is approximately conserved as it is strained. The consequence is that the total volume of the reference and deformed arteries are almost identical. In morphoelasticity, we assume that tissue first undergoes a stress-free volumetric growth (characterized by a growth tensor) before being strained. Specifically, the cuboid first responds to a pure growth meaning that each incremental part of the tissue increases its volume by a specified amount. This might cause an "overlap" in neighboring cuboids resulting in an incompatible configuration. Therefore by itself, the growth tensor does not give a physical configuration and needs to be followed by a corrective strain, characterized by an elastic tensor. The resulting compound process guarantees that the outcome is a continuous grown domain with no overlaps. Mathematically the process above assumes that the deformation gradient can be written as a product of the growth tensor ($\mathbf{F_g}$) and the elastic tensor ($\mathbf{F_e}$):

$$\mathbf{F} = \mathbf{F_e F_g}, \qquad (9)$$

see Figure 4. The energy in (7) also needs to be modified under this assumption of growth: for details, see (Fok & Gou, 2019).

How does morphoelasticity help with plaque modeling? The growth tensor $\mathbf{F_g}$ can be informed by the biology of atherosclerosis. From murine models the initial stages of plaque growth are characterized by an increase in intima mass, resulting from a migration of smooth muscle cells from the media (Clowes, et al., 1983). One possible trigger for this migration is PDGF (Jackson, et al., 1993). In our model, we simplify this process by assuming that PDGF is directly responsible for growth and take $\mathbf{F_g} = \begin{pmatrix} g_1 & 0 & 0 \\ 0 & g_2 & 0 \\ 0 & 0 & 1 \end{pmatrix}$: increments of arterial tissue increase their *radial* and *circumferential* dimensions by $g_1$ and $g_2$ respectively and we let $g_1$ and $g_2$ depend on PDGF concentration, (as predicted by equation (5)) through a Hill function

$$g_k = \exp\left( \alpha_k \int_0^t \frac{P^m}{P^m + P_0^m} dt' \right), \qquad (10)$$

where $\alpha_1$ = 1, $\alpha_2$ = 0.25, $m$ = 4 and $P_0$ = 1.12*10$^{-5}$ mol/L (these constants were chosen for convenience). The growth tensor is a crucial element of the model because it allows the histological and morphoelastic models to communicate with each other. The resulting *integrated* model of atherosclerosis allows us to understand how morphological features of plaque (e.g. lumen size, stenosis, necrotic fraction) could be affected by microbiology (e.g. oxLDL density, glucose concentrations, PDGF density) and can be summarized by the following processes:

- Oxidized LDLs infiltrate the intima and diffuse.
- Macrophages chemotax towards oxidized LDLs, consume them and release MCP1, attracting more macrophages.

- Macrophages die if oxygen levels are low, producing necrotic cells.
- PDGF released from platelets induces growth of neo-intima and a blood pressure acts on the arterial lumen. Both factors evolve plaque morphology over time.

Together, these four processes are able to produce a vast range of morphological and histological changes in the vessel. Our simulations are done in two dimensions since plaque structure is often presented in two-dimensional arterial cross sections. One possible realization of the model at a single time point is shown in Figure 5. We see that oxygen and macrophages are localized near the arterial lumen; a single localized injury to the endothelium has released PDGF, leading to elevated levels of the growth factor near the south pole of the lumen; and a large necrotic core has resulted from the death of many MCs deep in the intima.

**Discussion and Results.**

Now we illustrate the results of our model. Although there are many ways to produce different plaque outcomes, we focus on changing 5 parameters over time, reflecting 5 main mechanisms that are thought to be associated with inflamed plaques:

1- *Changes in oxLDL distribution, $L(t)$.* It has long been hypothesized that modified LDLs are atherogenic since they are mainly responsible for the appearance of the foam cell phenotype (Brown & Goldstein, 1983). The oxLDL density inside a plaque is governed by the function $L$ whose time evolution we control directly in the model.

2- *Changes in "resilience" of macrophages, $C_{\text{crit}}(t)$.* This parameter is a measurement of the resilience of macrophages to oxygen tension which (in our model) can be controlled over time. Biologically, macrophage resilience to hypoxic conditions is related to processes such as HIF regulation (Koh & Powis, 2012). Tougher macrophages result in less hypoxic death and possibly less necrosis.

3- *Changes in inflammation, $M_0(t)$.* We define inflammation in the endothelium as the density of macrophages adsorbed on the layer. Mathematically we control this density by a *boundary condition $M = M_0(t)$* for equation (2) for a given function $M_0(t)$ and solutions to equation (2) must satisfy this boundary condition. Biologically, an increasing $M_0(t)$ corresponds to an endothelium that becomes more inflamed over time.

4- *Development of vasa vasora, f(x,y,t)*. Vasa vasora are a network of small vessels that supply the arterial walls with resources such as oxygen and glucose. Sources of oxygen are controlled in our model through the function *f(x,y,t)* in equation (4).

5- *Changes in vessel pressure, p(t)*. The blood pressure from equation (8) affects lumen size and can increase in patients that develop hypertension.

A summary of these variables is provided in Table 2. In addition to these five mechanisms, intimal growth depends on PDGF (*P*) through eq. (10). Similar to mechanism **3-**, we control PDGF levels in the plaque by changing the boundary condition for *P*. Mathematically, we impose P = $P_0(x,t)$ for $x \in \partial\Omega$ with $P_0 > 0$ on $n$ = 1 or 2 small segments of $\partial\Omega$, representing discrete injury points on the endothelium, and $P_0 = 0$ otherwise. The number of injury points *n* is fixed in each case and controls how quickly the plaque grows (larger *n* corresponds to faster growth).

Now we make connections to the enhanced IVUS images from Kubo's paper. In this study, the authors imaged the same plaques 12 months apart to gain insight into their natural history (see Figure 1). There are 5 cases: A-E. To recreate these evolutions, the following protocols were used. Our parameters are separated into two types: those that can potentially change in time and reflect the 5 mechanisms given above, and those that remain static (but can differ from case to case). The quantities *L*, $C_{\text{crit}}$, $M_0$, *f* and *p* are in the first category (see Table 2) and all other parameters are in the second (see Table 1). The virtual plaques are evolved over a period of 12 months (up to *t* = 3 in simulation time) and tissue was defined as necrotic when *N* exceeded a threshold of 1.4 $\times$ $10^6$ cells/cm$^3$; otherwise the tissue was classified as fibrotic (see Figure 2). Other components (calcific, lipidic) were not directly accounted for by the model. Our simulation results are shown in Figure 6, which should be compared to Figure 1. Parameter values for each of cases A-E are given in Tables 1 and 2. Computer-generated animations of all five plaque evolutions can be found in the supplementary materials.

Case A results from a combination of mechanisms 2 and 3. We increase the macrophage resilience by decreasing $C_{\text{crit}}$ while also increasing the cap inflammation in time by increasing $M_0$. The result is plaque growth along with a necrotic core drifting deeper into the plaque, in rough qualitative agreement with the IVUS results in Figure 1A. Case B is achieved through mechanisms 2 and 5. We suppress PDGF completely so there is no endothelial injury or intimal growth and arterial morphology changes only because the lumen pressure *p* increases. MCs become more resilient as $C_{\text{crit}}$ is decreased. The overall result is that the intima thins, and the necrotic core vanishes.

Case C results from a combination of mechanisms 1 and 2. We model the movement of oxLDL density *L* to the right which attracts MCs. The necrotic core follows the macrophage density and also moves to the right because NCs are continuously produced by MC death. Case D results from a combination of mechanisms 2 and 4. Static oxygen sources model the effect of vasa vasora so that the baseline plaque is well-oxygenated. However we take MCs to become more susceptible to hypoxic death over time by increasing $C_{crit}$. The result is that as the plaque grows a necrotic core, consistent with the IVUS image in Figure 1D. Finally, we believe that changes in the *spatial distribution* of microvessels (mechanism 4) are key to producing the pair of images in Case E of Figure 1. There is a scattering of necrotic cells in the baseline state but they are highly dynamic and have increased in number in the follow-up case. The seemingly random spatial distribution of necrosis in the follow-up could result from changes in the number and density of vasa vasora in the intima, reflected by taking *f* to be a function that changes both temporally and spatially.

**Conclusions**

Plaques present considerable inter-patient variability, depending on location and age of the plaque, local and global cell biology, personal genetics, and environmental and lifestyle factors. It is not surprising that the associated evolutions seem hopelessly complicated. We advocate that integrated mathematical models may help with understanding how plaque complexity results from microbiology.

In this paper we assumed that equations (1)-(7) are universal, applying to all plaques (this assumption will be further discussed below) and different plaque states and evolutions are characterized by different parameters. Studying plaques through these equations is still daunting because there are 20 different parameters and functions in Tables 1 and 2. While in this paper we estimated them for the purposes of matching the IVUS images in Fig. 1, in principle each one should be measured *in-vivo* for a particular patient. Although this may not be practical or even possible, our basic understanding of plaque is still significantly enhanced by using this model as a framework, for two reasons. First, even though the parameters and driving functions are unknown, they could be estimated from *in-vitro* experiments or from animal models. Knowing more parameters gives the model more predictive power and puts the study of plaque on a more rigorous scientific footing. Second – and more importantly – the integrated model can be systematically tested and recalibrated to accommodate new data. Rather than passively describing plaques and accumulating more observations, the integrated model puts clinicians in the driving seat, enabling them to formulate and test new hypotheses. For example, in our results we saw that $C_{crit}$ (characterizing the sensitivity of macrophages to oxygen tension) was an important parameter in 4 out of the 5 cases. Our model quickly generated the hypothesis that necrosis fraction depends on $C_{crit}$. This hypothesis could, in principle, be tested through experiment by silencing the HIF-1$\alpha$ gene in macrophage cell lines (for example).

There is the remaining issue that equations (1)-(7) are probably *not* universal. However, this does not pose any added conceptual difficulty providing experimentalists, clinicians and theorists are willing to work closely together. Instead of updating parameter values, individual terms in the equations can be modified to reflect the latest data. New equations can even be included (or removed) if required. The power of a mathematical model is that it can be continuously refined and updated in a systematic way.

**Caveats and Limitations**

The model presented in this paper has focused on six main quantities: necrotic cells, macrophage cells, Monocyte Chemotactic Protein, Platelet-Derived Growth Factor, oxygen and oxidized LDLs. This is a minimal model: for example we have not distinguished between M1 and M2 macrophages (M1s are thought to be pro-inflammatory while M2s are atheroprotective), neither have we accounted for the effect of glucose or the intricate biology of cell death. However, the extension to these more complex cases is not conceptually more difficult in terms of PDE modeling. In principle one can write down as many equations as the biology demands. Physical laws realized through a mathematical framework provide systematic ways to do this that are consistent with principles of mass conservation and continuum mechanics. The main technical hurdle is not in the mathematical modeling but in finding parameter values and making comparisons with data. To do this requires careful calibration of the model output with serial images such as Figure 5. It also requires close collaboration between mathematicians/engineers and clinicians/imaging specialists.

Another limitation of the current model is that we have not accounted for residual stresses in the arterial segments at baseline (Figure 6, left column). When lumen pressure is removed from arterial segments *ex-vivo*, a stress called a residual stress still remains (this can be seen when a radial cut is made, usually causing the segment to spring open). The residual stress can alter the total Cauchy stress after pressurization and growth, so predictions of stress must be made very carefully with this model. Incorporating a residual stress can be done by modifying the reference configuration (Vandiver, 2014).

Other chemical messengers can be incorporated into the model depending on what is suggested by imaging data. Currently, enhanced IVUS provides spatially resolved maps of necrosis, cholesterol and calcium phosphate *in-vivo*. While other cytokines such as interleukins (Mallat, et al., 1999; Huber, et al., 1999) also play an important role in the development of plaques, including them into our model does not provide additional insight if there are no data to indicate their spatial distribution and temporal evolution; in fact including their effect would only increase the number of unknown parameters in our model and therefore the uncertainty of the predictions. Histological stains such as Movat and Haematoxylin & Eosin (H&E) constitute the "gold standard" in terms of determining spatial data within the plaque. If these stains can be used in an animal study in which animals are periodically sacrificed and particular arteries analyzed in terms of cell positions and cell types, this experimental protocol could provide valuable information about the dynamics within a plaque and further inform our integrated model.

**Future Challenges**

The future is promising. Imaging technologies such as OCT and infrared spectroscopy are giving us the ability to visualize the evolution of plaque morphology and composition. The great challenge at the moment is the way the data are being collected. Scientific prediction is concerned with quantifying future states from past states. The more informed we are of past states, the better our prediction of future states. The few existing "natural history" studies of human atherosclerosis have an extremely low time resolution. For example, the (Kubo, et al., 2010) plaques are scanned at just two time instances (a baseline and a follow-up). Serial studies in single patients need to be performed more frequently, and at greater time resolution. The hurdle here is clinical indication for the procedure. Nevertheless, we believe that high time-resolution serial studies hold the key to a better understanding of atherosclerosis. Providing modelers and quantitative scientists with more snapshots, or even a movie of disease progression will greatly enhance the modeling and possibly prediction of atherosclerotic development.

Another advance that will greatly help in aiding prediction is an improvement in virtual histology technology and integration with other imaging methods. Currently the master algorithm classifies tissues into four categories: fibrous, fatty, necrotic and calcific. In reality, a given region can share characteristics from each category. For example, necrotic cells could be interspersed with flecks of calcium. What is really needed is a continuous version of the 4 categories. What could such a scale look like? One possibility is to use a vector that represents weights of each category: p = ($p_1$, $p_2$, $p_3$, $p_4$) where $\sum_{i=1}^{4} p_i = 1$. For example, (0,0,0,1) corresponds to a region that is completely calcified and (0,0,1,0) corresponds to a region that is completely necrotic. However, an area where calcium and necrosis are present in equal amounts can be represented as (0, 0, 0.5, 0.5). Constructing the algorithms to provide this level of detail will prove challenging, since even the current 4-color method is not universally accepted and does not always match the histological gold-standard.

Overall, the mathematical community is well-positioned to make inroads to undetstanding the patterns seen in enhanced IVUS images or other imaging modalities of atherosclerotic plaque. Fundamentally, necrotic cores form because cells in the plaque die and are not cleared away in the usual way by phagocytes. An increase in cell number within the plaque must come from either proliferation or cell fluxes. While the underlying biology in plaque development is intricate and complex, we believe that mathematics provides the conceptual bridge that connects plaque environment to the spatial patterns that ultimately result.

## Bibliography

Akyildiz, A. C. et al., 2011. Effects of intima stiffness and plaque morphology on peak cap stress. *Biomedical engineering online,* p. 10 (25).

Auricchio, F., Di Loreto, M. & Sacco, E., 2001. Finite-element analysis of a stenotic artery revascularization through a stent insertion.. *Computer Methods in Biomechanics and Biomedical Engineering,* pp. 249 - 263.

Bendeck, M. et al., 1994. Smooth muscle cell migration and matrix metalloproteinase expression after arterial injury in the rat. *Circulation research,* pp. 539-545.

Benjamin, E., Muntner, P. & Bittencourt, M., 2019. Heart Disease and Stroke Statistics—2019 Update: A Report From the American Heart Association. *Circulation.*

Bourantas, C. V. et al., 2013. Clinical and angiographic characteristics of patients likely to have vulnerable plaques: analysis from the PROSPECT study. *JACC: Cardiovascular Imaging,* pp. 1263-1272.

Brown, M. & Goldstein, J., 1983. Lipoprotein metabolism in the macrophage: implications for cholesterol deposition in atherosclerosis.. *Annual review of biochemistry,* 52(1), pp. 223 - 261.

Clowes, A. W., Reidy, M. A. & Clowes, M. M., 1983. Kinetics of cellular proliferation after arterial injury. I. Smooth muscle growth in the absence of endothelium. *Laboratory investigation; a journal of technical methods and pathology,* pp. 327-333.

Cyrus, T. et al., 1999. Disruption of the 12/15-lipoxygenase gene diminishes atherosclerosis in apo E–deficient mice.. *The Journal of clinical investigation,* pp. 1597-1604.

DiMilla, P., Barbee, K. & Lauffenburger, D., 1991. Mathematical model for the effects of adhesion and mechanics on cell migration speed.. *Biophysical Journal,* pp. 15-37.

Fok, P.-W., 2012. Growth of Necrotic Cores in Atherosclerotic Plaque. *Mathematical Medicine and Biology,* 29(4), pp. 301-327.

Fok, P. & Gou, K., 2019. Finite Element Simulation of Intimal Thickening in Multi-Layered Arterial Cross Sections by Morphoelasticity. *Computer Methods in Applied Mechanics and Engineering (under review).*

Geer, J. C., McGill, H. C. & Strong, J. P., 1961. The Fine Structure of Human Atherosclerotic Lesions. *The American Journal of Pathology,* pp. 263-287.

Goriely, A., 2017. *The mathematics and mechanics of biological growth.* New York: Springer.

Holzapfel, G. A., 2000. *Nonlinear Solid Mechanics: A Continuum Approach for Engineering.* Chichester: John Wiley and Sons.

Holzapfel, G., Sommer, G., Gasser, C. & Regitnig, P., 2005. Determination of layer-specific mechanical properties of human coronary arteries with nonatherosclerotic intimal thickening and related constitutive modeling. *American Journal of Physiology - Heart and Circulatory Physiology,* pp. H2048 - H2058.

Huber, S. et al., 1999. Interleukin-6 exacerbates early atherosclerosis in mice. *Arteriosclerosis, thrombosis, and vascular biology,* pp. 2364 - 2367.

Jackson, C. L., Raines, E. W., Ross, R. & Reidy, M. A., 1993. Role of endogenous platelet-derived growth factor in arterial smooth muscle cell migration after balloon catheter injury.. *Arteriosclerosis and thrombosis: a journal of vascular biology,* pp. 1218 - 1226.

Koh, M. Y. & Powis, G., 2012. Passing the baton: the HIF switch. *Trends in Biochemical Sciences,* 37(9), pp. 364 -- 372.

Kubo, T. et al., 2010. The Dynamic Nature of Coronary Artery Lesion Morphology Assessed by Serial Virtual Histology Intravascular Ultrasound Tissue Characterization. *Journal of the American College of Cardiology.*

Loree, H. M., Kamm, R. D., Stringfellow, R. G. & Lee, R. T., 1992. Effects of fibrous cap thickness on peak circumferential stress in model atherosclerotic vessels. *Circulation Research,* pp. 850 - 858.

Mallat, Z. et al., 1999. Protective role of interleukin-10 in atherosclerosis. *Circulation research,* pp. e17 - e24.

Nair, A., Kuban, B. D., Obuchowski, N. & Vince, D. G., 2001. Assessing spectral algorithms to predict atherosclerotic plaque composition with normalized and raw intravascular ultrasound data. *Ultrasound in medicine and biology,* 10(29), pp. 1319-1331.

R, V., AP, B., FD, K. & A, F., 2002. Vulnerable plaque: the pathology of unstable coronary lesions. *Journal of interventional cardiology,* 15(6), pp. 439-446.

Simon, B., Kaufmann, M., McAfee, M. & Baldwin, A., 1993. Finite element models for arterial wall mechanics.. *Journal of biomechanical engineering,* pp. 489 - 496.

Stadius, M. L. et al., 1992. Time course and cellular characteristics of the iliac artery response to acute balloon injury. An angiographic, morphometric, and immunocytochemical analysis in the cholesterol-fed New Zealand white rabbit. *Arteriosclerosis and thrombosis: a journal of vascular biology,* pp. 1267 - 1273.

Vandiver, R., 2014. Effect of residual stress on peak cap stress in arteries. *Mathematical Biosciences and Engineering,* 11(5), pp. 1199 - 1214.

Vengrenyuk, Y. et al., 2006. A hypothesis for vulnerable plaque rupture due to stress-induced debonding around cellular microcalcifications in thin fibrous caps. *Proceedings of the National Academies of Science of the United States of America,* 103(40).

Virmani, R., Burke, A. P., Kolodgie, F. D. & Farb, A., 2003. Pathology of the Thin-Cap Fibroatheroma. *Journal of Interventional Cardiology,* pp. 267 - 272.

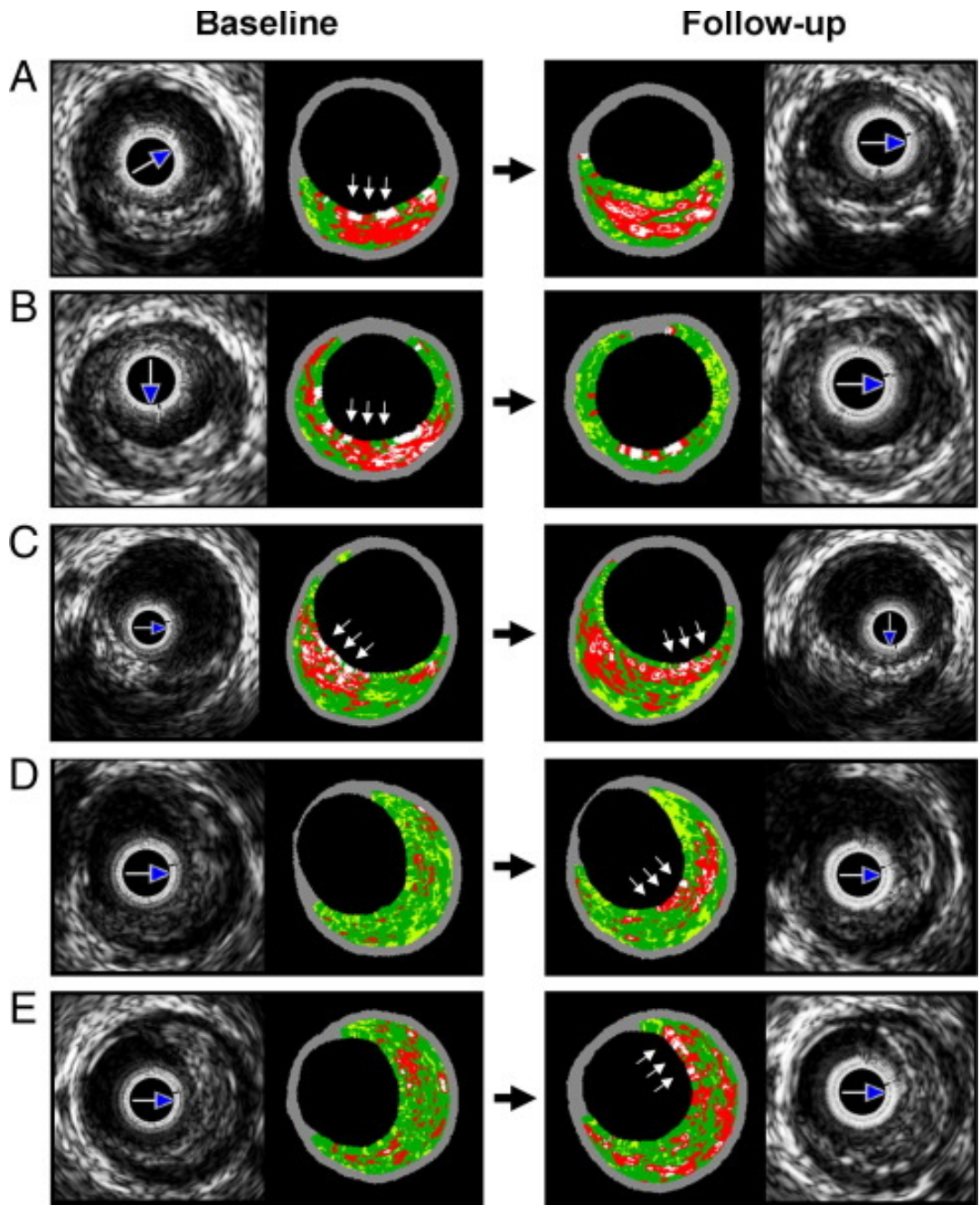

Figure 1: Serial images from (Kubo, et al., 2010). Baseline and follow-up were 12 months apart.

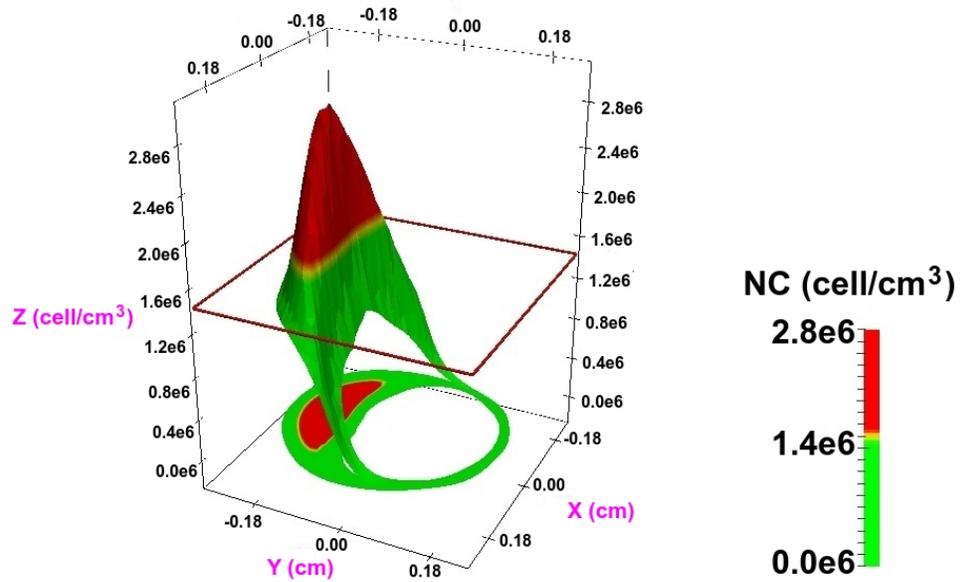

**Figure 2:** Solving the PDEs (1)-(5) produces smooth functions for biological quantities such as the necrotic cell density, N(x,y,t). By introducing a threshold value (here N = $1.4*10^6$), the function can be used to explain regions of necrosis in advanced fibroatheromas: $N<1.4*10^6$ indicates the presence of fibrotic tissue (green) while $N>1.4*10^6$ indicates the presence of necrotic tissue (red). The 3D surface N(x,y,t) can be projected onto the XY plane to recreate 2D IVUS cross sections in Fig. 1.

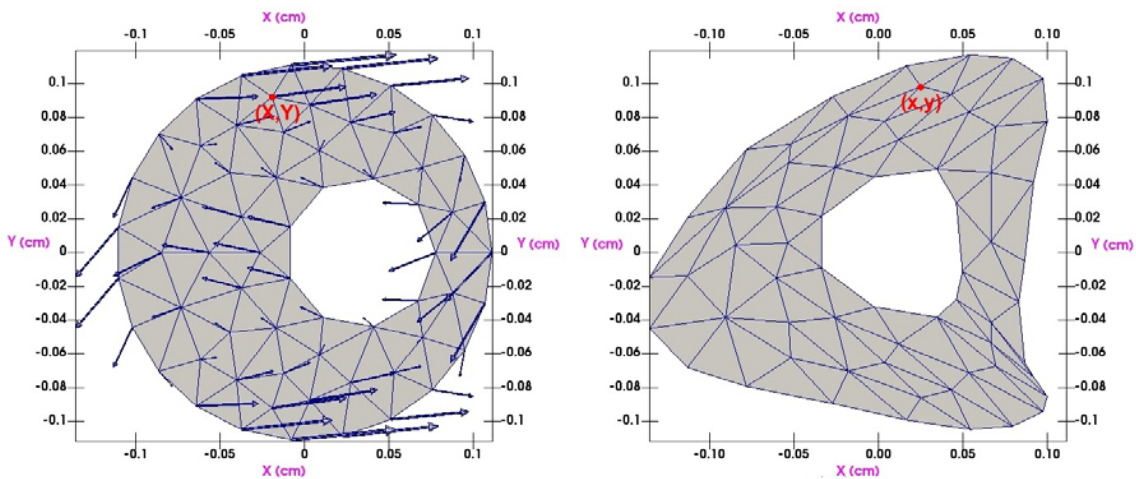

**Figure 3:** Minizing the energy (7) produces a displacement field (indicated by arrows) that deforms the arterial cross section.

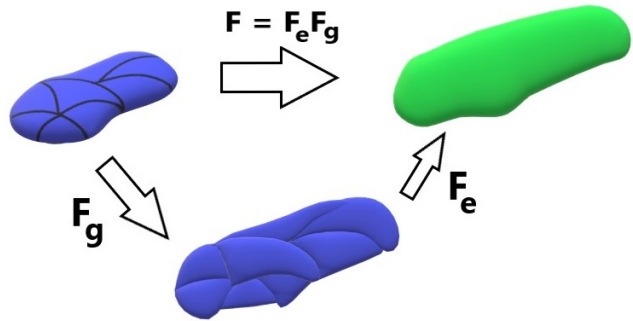

Figure 1: Decomposition of the deformation gradient F into a pure growth and an elastic response. This is the fundamental assumption in morphoelasticity theory: see Eq. (9).

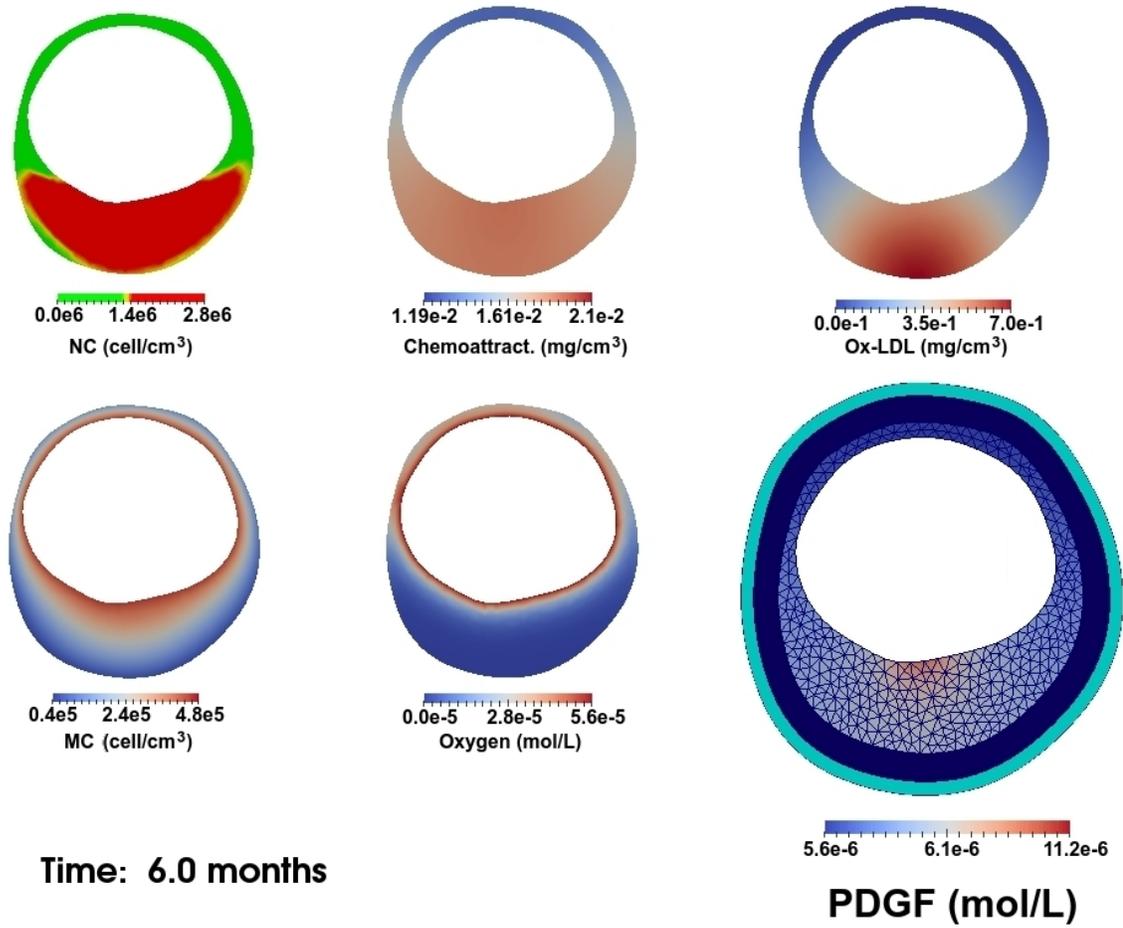

Figure 2: Typical plaque model results. Concentrations of NC, Chemoattractant, oxLDL, MC and and oxygen are outputted as spatially dependent fields distributed within the intima. For PDGF concentration, the media and adventitia are also indicated along with the finite element mesh.

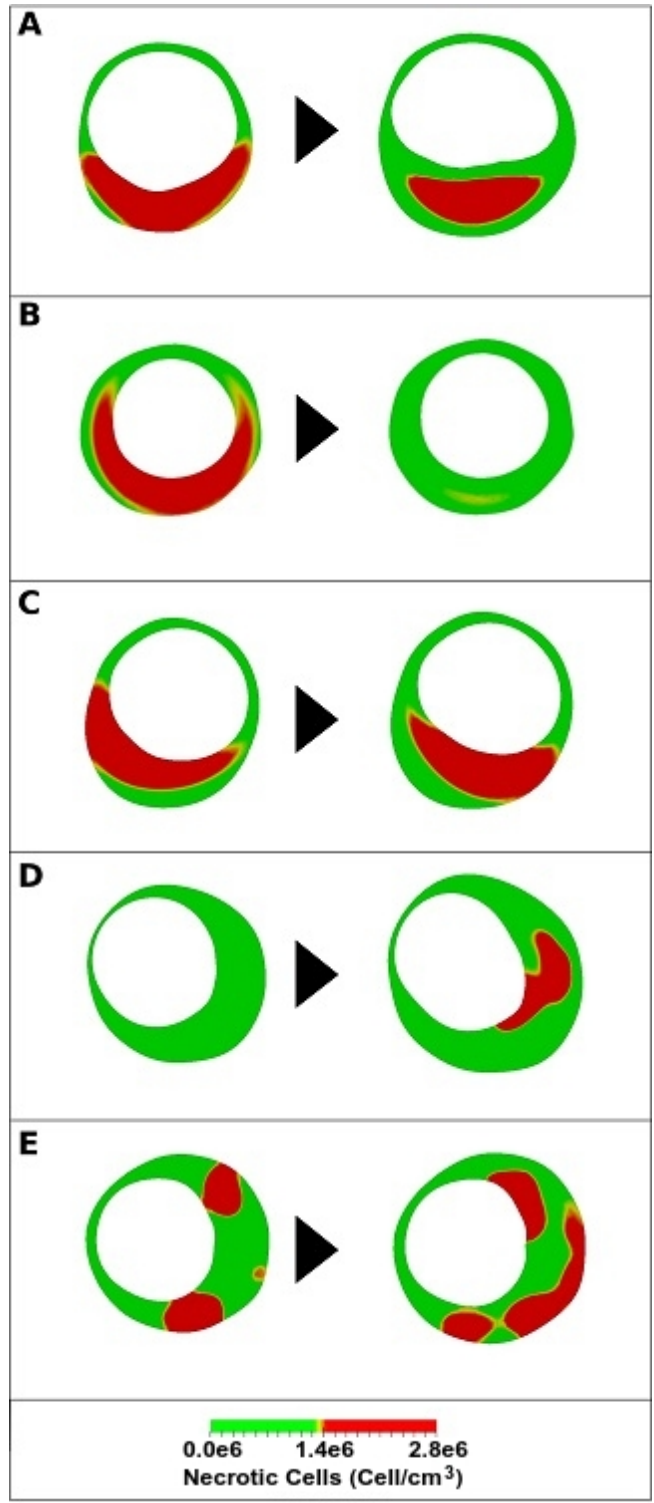

Figure 6: Simulations of necrotic core development in a coronary artery (compare with Fig. 1)

| Symbol | Meaning | Case A | Case B | Case C | Case D | Case E | Units |
|---|---|---|---|---|---|---|---|
| $\mu$ | Chemotactic coefficient of MCs | 2.26 | 7.77 | 4.41 | 9.17 | 6.02 | (mm²/day).(cm³/mg) |
| $D_1$ | Diffusivity of NCs | $3.2 \times 10^{-3}$ | $3.3 \times 10^{-3}$ | $3.6 \times 10^{-3}$ | $3.9 \times 10^{-3}$ | $4.3 \times 10^{-3}$ | cm²/day |
| $D_2$ | Diffusivity of MCs | 1.72 | 1.74 | 1.92 | 2.06 | 2.30 | cm²/day |
| $D_3$ | Diffusivity of CKs | 80.8 | 81.6 | 89.7 | 96.3 | 107.6 | cm²/day |
| $D_4$ | Diffusivity of O₂ | 172.5 | 174.3 | 191.6 | 205.8 | 229.9 | cm²/day |
| $D_5$ | Diffusivity of PDGF | 19.4 | 0 | 32.3 | 15.4 | 21.5 | cm²/day |
| $\beta_1$ | Clearance rate of NCs | 0.1 | 0.1 | 0.1 | 0.1 | 0.1 | day⁻¹ |
| $\beta_2$ | Clearance rate of CKs | 10 | 10 | 10 | 10 | 10 | day⁻¹ |
| $\beta_3$ | Background O₂ consumption rate | $1.2 \times 10^4$ | $1.2 \times 10^4$ | $1.2 \times 10^4$ | $1.2 \times 10^4$ | $1.2 \times 10^4$ | day⁻¹ |
| $\beta_4$ | Decay rate of PDGF | 2 | 0 | 2 | 2 | 2 | day⁻¹ |
| $\lambda_1$ | Production rate of CKs | $2.5 \times 10^{-6}$ | $2.5 \times 10^{-6}$ | $2.5 \times 10^{-6}$ | $2.5 \times 10^{-6}$ | $2.5 \times 10^{-6}$ | cm³/day |
| $\lambda_2$ | Consumption rate of O₂ by MCs | $2.5 \times 10^{-7}$ | $2.5 \times 10^{-7}$ | $2.5 \times 10^{-7}$ | $2.5 \times 10^{-3}$ | $2.5 \times 10^{-7}$ | cm³/day |
| $\gamma_{min}$ | Normoxic MC death rate | $3 \times 10^{-3}$ | $3 \times 10^{-3}$ | $3 \times 10^{-3}$ | $3 \times 10^{-3}$ | $3 \times 10^{-3}$ | 1/day |
| $\gamma_{max}$ | Hypoxic MC death rate | 1.2 | 1.2 | 1.2 | 1.2 | 1.2 | 1/day |
| $n$ | # Injury Points | 1 | 0 | 1 | 2 | 1 | none |

Table 1: Parameter values for Eqs (1)-(5) used to generate Figure 6. Abbreviations: MC = Macrophage Cells, NC = Necrotic Cells, CKs = Chemokines, O₂ = Oxygen, PDGF = Platelet Derived Growth Factor.

Table 2: Time-dependent parameters for Mechanisms 1-5.

| | Meaning | Case A | Case B | Case C | Case D | Case E | Units |
|---|---|---|---|---|---|---|---|
| $L(t)$ | OxLDL concentration | Static | Static | Dynamic | Static | Static | mg/cm$^3$ |
| $C_{crit}(t)$ | Normoxic-Hypoxic O$_2$ threshold | Decreases from $2.80 \times 10^{-5}$ to $0.56 \times 10^{-5}$ | Decreases from $2.80 \times 10^{-5}$ to $2.24 \times 10^{-5}$ | Decreases from $2.80 \times 10^{-5}$ to $2.04 \times 10^{-5}$ | Increases from $1.68 \times 10^{-5}$ to $5.04 \times 10^{-5}$ | $3.64 \times 10^{-5}$ | mol/L |
| $M_0(t)$ | Macrophage density in endothelium | Increases from $4 \times 10^5$ to $4.2 \times 10^5$ | $4 \times 10^5$ | $4 \times 10^5$ | $4 \times 10^5$ | $4 \times 10^5$ | cells/cm$^3$ |
| $f(x,y,t)$ | O$_2$ sources within intima | 0 | 0 | 0 | Sum of Gaussian functions[1] | Sum of Gaussian functions[2] | mol/L/day |
| $p(t)$ | Lumen pressure | 90 | Increases from 90 to 143 | 90 | 90 | 90 | mmHg |

---

[1] Specifically, $f(x,y,t) = f(x,y) = \sum_{k=1}^{N} f_k \exp[-a_k(x-x_k)^2 - b_k(y-y_k)^2]$ for some time-independent constants $f_k$, $a_k$, $b_k$ and $N$ and Gaussian centers $(x_k, y_k)$.

[2] Specifically, $f(x,y,t) = \sum_{k=1}^{N} f_k(t) \exp[-a_k(x-x_k)^2 - b_k(y-y_k)^2]$ for some constants $a_k$, $b_k$, $N$, $x_k$, $y_k$ and time-dependent functions $f_k(t)$.